\documentclass[letterpaper]{article}

\pdfoutput=1
\usepackage[final]{nips_2016}
\usepackage[utf8]{inputenc} 
\usepackage[T1]{fontenc}    
\usepackage{hyperref}       
\usepackage{url}            
\usepackage{booktabs}       
\usepackage{amsfonts}       
\usepackage{nicefrac}       
\usepackage{microtype}      

\usepackage{graphicx}
\usepackage{bm}
\usepackage{siunitx}
\usepackage{appendix}
\usepackage{multirow}
\usepackage{afterpage}

\title{A Bayesian Approach to Predicting Disengaged Youth}

\author{
  David Kohn \\
  Centre for Translational Data Science \\
  University of Sydney \\
  New South Wales 2006 \\
  \texttt{david.kohn@sydney.edu.au} \\
  \And
  Sally Cripps \\
  Centre for Translational Data Science \\
  University of Sydney \\
  New South Wales 2006 \\
  \And
  Nick Glozier \\
  Brain Mind Centre \\
  University of Sydney \\
  New South Wales 2006 \\
  \And
  Hugh Durrant-Whyte \\
  Centre for Translational Data Science \\
  University of Sydney \\
  New South Wales 2006 \\
}

\begin{document}

\maketitle

\begin{abstract}
This article presents a Bayesian approach for predicting and identifying the factors which most influence an individual's propensity to fall into the category of Not in Employment Education or Training (NEET).
The approach partitions the covariates into two groups: those which have the potential to be changed as a result of an intervention strategy and those which must be controlled for.
This partition allows us to develop models and identify important factors conditional on the control covariates, which is useful for clinicians and policy makers who wish to identify potential intervention strategies.
Using the data obtained by \citet{ODea2014} we compare the results from this approach with the results from \citet{ODea2014} and with the results obtained using the Bayesian variable selection procedure of \citet{Lamnisos2009} when the covariates are not partitioned.
We find that  the relative importance of predictive factors varies greatly depending upon the control covariates.
This has enormous implications when deciding on what interventions are most useful to prevent young people from being NEET.
\end{abstract}

\section{Background}
Often the number of factors available for prediction on a given individual, $p$, is larger than the number of individuals on whom we have NEET status measurements, $n$, making the identification of important factors statistically challenging.
For example, inference in a frequentist procedure usually relies on the assumption of asymptotic normality of the sample estimates, however as $p\rightarrow n$  this assumption is unlikely to be true.
Another related issue is that good predictive performance does not necessarily equate with the identification of causal factors; many different combinations of factors may be equally good in predicting whether or not an individual will end up in the NEET category.
However, if issues such as NEET are to be addressed, policy makers need to know what modifiable factors are likely to be causal so that appropriate intervention strategies are used.

To address the joint issues of high dimensionality and causal inference we take a Bayesian approach.
Specifically, we propose to reduce the dimensionality by using a spike and slab prior over the regression coefficients \citep[see, for example,][]{Lamnisos2009}.
This regularization may result in biased estimates of the regression coefficients as per \citet{Chernozhukov2016}.
To address this issue we develop a series of conditional models by dividing the covariates into two groups, those which have the potential to be changed by an intervention, for example an individual's clinically assessed depression score,  and those which do not, for example an individual's age or sex.
These conditional models are typically based on very small sample sizes, with $p > n$, making variable selection difficult but important.

\section{Model}
Suppose we have observations on $n$ individuals' NEET status, $\bm y=(y_1,\ldots,y_n)$, where $y_i=1$ if an individual $i$ is classified as NEET and $y_i=0$ otherwise and corresponding measurements on  covariates for each of these individuals.
 We denote the potentially causal factors by $X$ and the other factors, which we refer to as control factors, by $W$, with $X=(1,\bm x_1',\ldots, \bm x_n')$, where $\bm x_i=(x_{i1},\ldots,,x_{iP_x} )$, and $x_{ki}$  the measurement of the $k^{th}$ covariate on individual $i$, for $i=1,\ldots,n$, and $k=1,\ldots,P_x$, and $P_x$ is the number of covariates which are potentially modifiable.  
We model the dependence between $y$, conditional on the control variables, $w$, and $x$ using a generalized linear model (GLM),
\begin{equation}
\Pr\left(y_{i}=1|\bm w_i,\bm x_{i}\right)=g\left(\bm x_{i}\bm\beta(\bm w_i)\right),
\label{eq_bin_reg}
\end{equation}
where $g$ is some link function, and the notation $\bm\beta(\bm w)$ means that the $(P_x+1) \times 1$ vector of regression coefficients is parameterized to depend  upon the control variables $W$.
This paper uses the standard normal CDF as the link function, so that $g(X\bm\beta(w))=\Phi(X\bm\beta(w))$, where $\Phi(X\bm\beta(w))=\Pr(z<X\bm\beta(w))$ and $z\sim N(0,1)$.
The choice of $\Phi$ as the link function allows us to employ the data augmentation method of \citet{Albert1993} to estimate the regression coefficients and perform variable selection.

To obtain the likelihood function we divide the data into $S \le n$ non-overlapping partitions,  $W=(\bm w_1,\ldots,\bm w_S)$, where $\bm w_s=(1,w_{s1},\ldots,w_{sP_w})$ represents a $(P_w+1) \times 1 $ vector of unique values of the control factors, for $s=1,\ldots S$, with $P_w$ the number of control covariates.
Let $n_s\ge1$, be the number of observations in each partition $s=1,\ldots, S$ and define $I_{s}$ to be the set of indices for observations corresponding to partition $s$ for $s=1,\ldots S$.
Then, the likelihood function is
\begin{equation}
p(\bm y|W,X,\bm\beta)=\prod_{s=1}^{S}\prod_{i\in I_s} \Phi(\bm x_i\bm\beta(\bm w_s))^{ y_{i}}(1-\Phi(\bm x_i\bm\beta(\bm w_s)))^{(1-y_{i})}.
\label{eq_like1}
\end{equation}
To fully specify the model we  place priors on those parameters needed to evaluate the likelihood, namely the regression coefficients.
We wish to place a prior on these regression coefficients to allow for the possibility that some causal factors on which we have measurements do not have an  impact on the future NEET status of an individual and  also to allow this possibility to depend upon the control factors.

Specifically, for each partition $s$, we introduce an indicator vector $\bm\gamma_s=(\gamma_{1s},\ldots,\gamma_{P_x,s})$, where  $\gamma_{ks}=1$ if causal factor $k$ is in the model for partition $s$ and $\gamma_{ks}=0$ otherwise.
To write the prior for the regression coefficients, we define the set $A_{1s}=\{k:\gamma_{ks}=1\}$ to be the set of indices corresponding to those causal factors which are included in the model for partition $s$  and define $A_{0s}$ similarly.

Finally, we define $ \Pr \left( \gamma_{ks} = 1 \right| \bm w_s)=\pi_k(\bm w_s) $  to be the prior probability that the $k^{th}$ causal factor is in the model for partition $s$.
This  is parameterized to depend on the control covariates $\bm w_s$.
We model the dependence between $\gamma_{k}$ and $\bm w_s$ as a probit regression, so
 \begin{equation}\pi_k(\bm w_s)=\Phi(\bm f_k(\bm w_s)),
\label{eqn_mix_fun}
\end{equation}
where $f$ is some, possibly non-linear, function.
For now we take $f$ as linear so that $f_k(\bm w_s)=\bm w_s \bm\alpha_k$, where $\bm\alpha_k=(\alpha_{k0},\ldots,\alpha_{kP_w})$, is the $(P_w+1)\times 1$ vector of regression  coefficients for (\ref{eqn_mix_fun}).

\section{Data Description}
\label{sec:data_desc}
We use data from the `Transitions Study' as used in \citet{ODea2014} and detailed in \citet{Purcell2015}.
The study was conducted at two time periods, `baseline' and `followup' and implemented at four mental health service centres in Australia.
The 377 participants in the sample are aged 16-25 and were subject to a variety of demographic questions and clinical and psychological assessments.
Our target variable is NEET status, defined as not being in employment, education or training in the past month, measured at both baseline and follow up periods.
Our other variables are measured only at the baseline period and include QIDS, which assesses the presence of major diagnostic symptoms of depression; WHO-ASSIST, which assesses risky use of tobacco, alcohol and cannabis; GAD, which measures the symptoms of generalized anxiety disorder; WHODAS, which is a self rated examination of perceived functioning in daily life; and the demographic factors age and sex.
To make interpretation easier we refer to QIDS, GAD and WHODAS as depression, anxiety and functioning respectively.

\section{Results}
In this analysis we consider two settings for the response variable, (i) an individual's NEET status at baseline and (ii) an individual's NEET status at follow up, and two settings for the covariates (i)  using depression, tobacco, alcohol, cannabis, anxiety, functioning, age and sex as potential predictive factors (ii) using only the "modifiable" factors depression, tobacco, alcohol, cannabis, anxiety, functioning as potential predictors. 
To compare our method with other techniques  we first analyse the data using commonly used Bayesian and frequentist techniques. 

For the Bayesian variable selection analysis we follow \citet{Lamnisos2009} and use a \textit{g}-prior prior over the regression coefficients. 
The results of this analysis appear in Table \ref{tab:bvs_no_partition}. Panel~(a) shows the results using baseline NEET as the target variable while panel~(b) uses followup NEET as the target variable. 
The quantity $\hat{\beta}$ is estimate of the posterior mean of $\beta$ conditional on $\beta\ne0$, i.e. $E(\beta|\bm y, \beta\ne 0)$.  The MPP label refers to the marginal posterior probability of inclusion for a variable.

Table \ref{tab:bvs_no_partition} shows that the covariates generally have higher MPPs for the baseline NEET analysis relative to the followup NEET analysis, indicating that the concurrent measurements are correlated with NEET status but are less relevant as temporal, causal factors. 
Sex and age have high MPPs in the baseline model, motivating our conditioning on these variables as control factors so we can get a more finely gridded estimate of the probability of inclusion for the modifiable  factors. 
The accuracy of the models, as given by the  area under the ROC curve, (AUC), estimated by 5 fold crossvalidation and seen in the bottom row of table \ref{tab:bvs_no_partition}, is similar.
5-fold crossvalidation on the average model for both baseline and followup results in relatively similar ROC area-under-curve (AUC) for all models, as seen in the bottom row of Table \ref{tab:bvs_no_partition}.

For the frequentist analysis we achieve sparsity in the regression model using two methods; (i) standard  stepwise regression (ii) L1 regularization penalty, Lasso. The results of these analyses appear in Table~\ref{tab:stepwise}.
The stepwise and Bayesian variable selection methods select  identical  variables for $y$=baseline NEET status. 
In stark contrast, the results from the Lasso analysis shows that the only variable with a p-value less than the traditional cut-off of 0.05, is sex.

\begin{figure}[!htbp]
  \centering
  \begin{minipage}[b]{0.4\textwidth}
    \includegraphics[width=\textwidth]{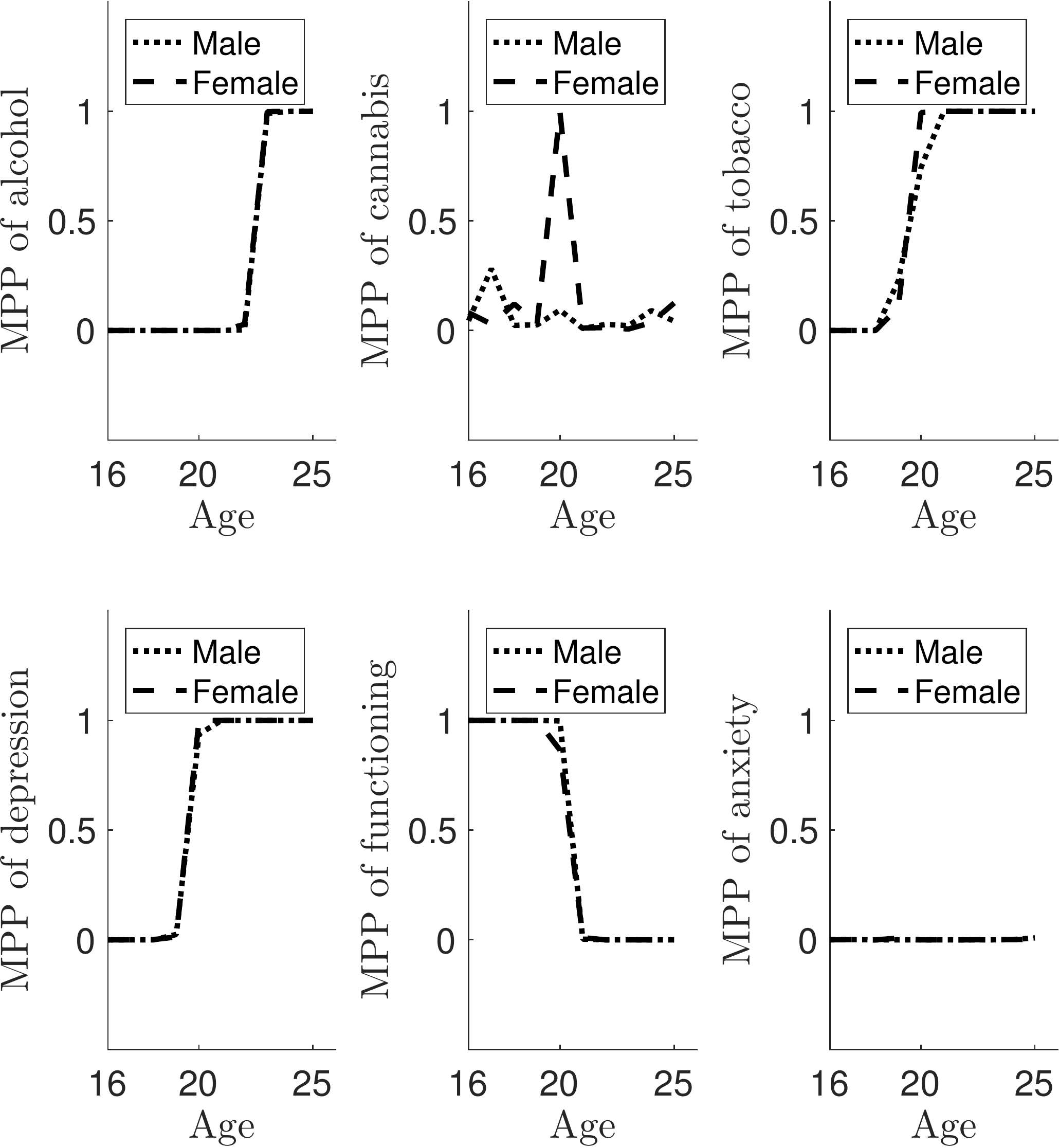}
  \caption{Marginal posterior probabilities (MPP) over the different partitions of age and sex for each modifiable variable  with $y= \text{baseline NEET status}$}
  \label{fig:mpp_baseline}
  \end{minipage}
  \hfill
  \begin{minipage}[b]{0.4\textwidth}
    \includegraphics[width=\textwidth]{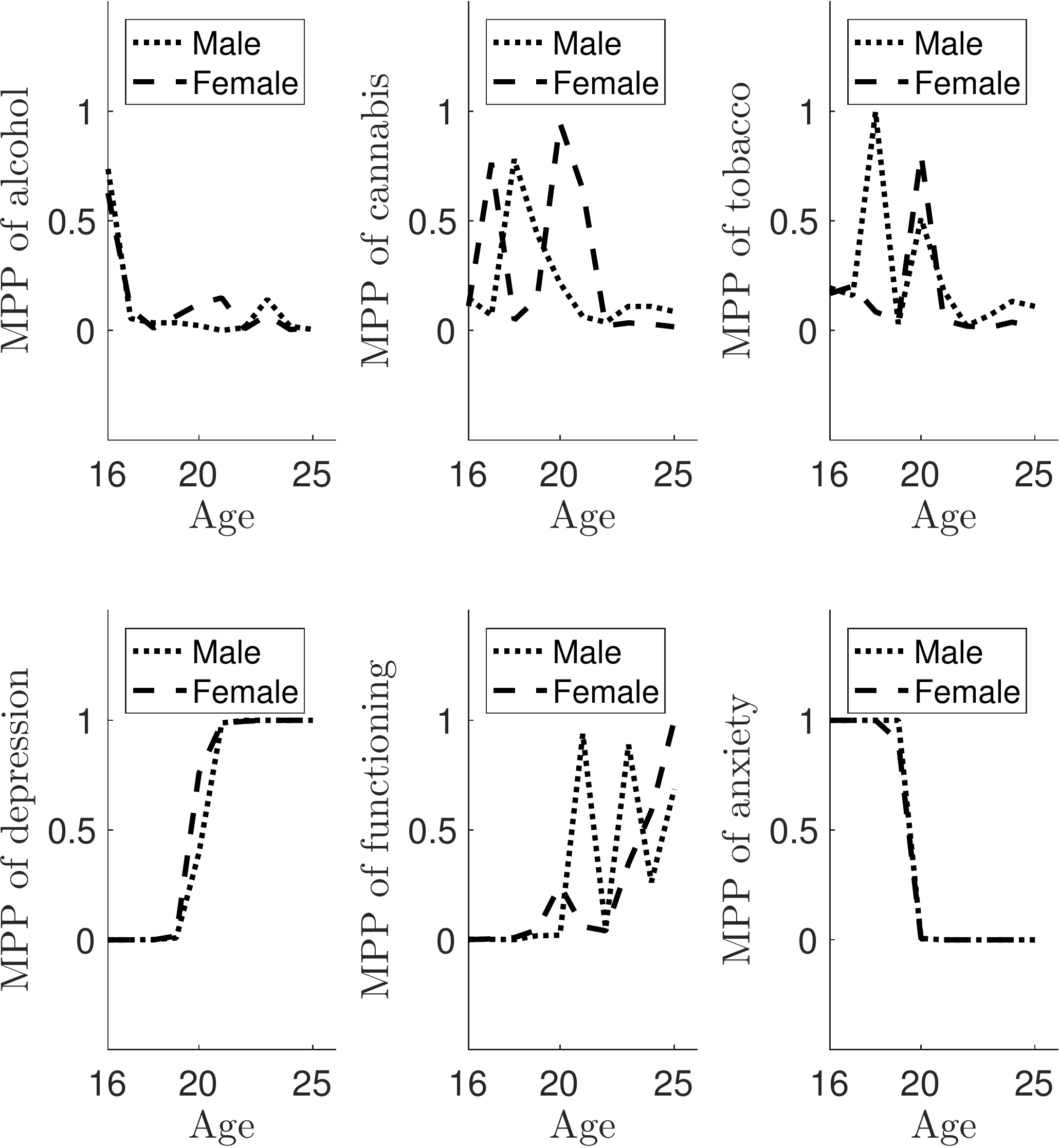}
  \caption{Marginal posterior probabilities (MPP) over the different partitions of age and sex for each modifiable variable with $y= \text{followup NEET status}$}
  \label{fig:mpp_followup}
  \end{minipage}
\end{figure}

The results of our partition method appear in Figures~\ref{fig:mpp_baseline}~and~ \ref{fig:mpp_followup}.   Figures~\ref{fig:mpp_baseline}~and~ \ref{fig:mpp_followup} show the posterior probabilities $\Pr(\gamma_{ks}=1|\bm y, \bm w)$, with $\bm w=(1, \mbox{Age}_s,G_s)$, where $\mbox{Age}_s$ is age years, ranging from 16 to 25 and $G_s=1$ corresponds to female and $G_s=0$ corresponds to male..
The number of distinct control categories, $S=20$.
Figures \ref{fig:mpp_baseline} and \ref{fig:mpp_followup} shows that the MPPs of the modifiable variables depends upon the control variables. In Fgure \ref{fig:mpp_baseline}, we see that the MPPs of alcohol, tobacco and depression all show a step-like increase from the ages of 19-21 onwards; that the MPPs of cannabis and tobacco spike at a few discrete ages; and that the MPP of functioning is high up to age 20 then has a step like decrease afterwards.
In Figure \ref{fig:mpp_followup}, we see that the MPPs of alcohol, cannabis, tobacco and functioning all show MPP spikes at various discrete ages depending on sex; that depression again shows a step-like increase in MPP from age 20 onwards; and that anxiety shows a step-like decrease in MPP from age 20 onwards.
The specific variables in the partition models are generally similar to those selected in the non-partition models which include the demographic variables.
However, we note that a lot of the `promising' variables are activated or have high MPPs at specific ages.
The effects of age generally make intuitive sense, the MPPs of alcohol and tobacco having a more probable relationship with NEET as the age of participants increase.
This is also confirmed by including interaction terms on the control covariates in the non-partition model, the results of which are consistent with the findings in the partition model.

Our method is able to provide more specific inferences about important variables within a given partition  and we see that our new  formulation of priors  provides meaningful differences in the interpretation of results.
 The results highlight the need to consider different intervention strategies for individuals with different non-modifiable factors.

\begin{table}[!htbp]
  \caption{Bayesian variable selection with no covariate partition}
  \label{tab:bvs_no_partition}
  \centering
  \begin{tabular}{l S[table-format=1.2] S[table-format=1.2] S[table-format=1.2] S[table-format=1.2] S[table-format=1.2] S[table-format=1.2] S[table-format=1.2] S[table-format=1.2]}
    \toprule
    & \multicolumn{4}{c}{(a) $y=\text{baseline NEET status}$} & \multicolumn{4}{c}{(b) $y=\text{follow up NEET status}$} \\
    \cmidrule{2-9}
    & \multicolumn{2}{c}{(i) Ex. age, sex} & \multicolumn{2}{c}{(ii) Inc. age, sex} & \multicolumn{2}{c}{(i) Ex. age, sex} & \multicolumn{2}{c}{(ii) Inc. age, sex}  \\
    \cmidrule{2-9}
    Variable & {$\hat{\beta}$} & {MPP} & {$\hat{\beta}$} & {MPP} & {$\hat{\beta}$} & {MPP} & {$\hat{\beta}$} & {MPP}  \\
    \midrule
    Alcohol & -0.01 & 0.16 & -0.03 & 0.97 & -0.00 & 0.08 & -0.00 & 0.17  \\
    Cannabis & 0.02 & 0.98 & 0.02 & 0.97 & 0.01 & 0.57 & 0.01 & 0.48 \\
    Tobacco  & -0.00 & 0.02 & 0.00 & 0.02 & 0.00 & 0.03 & 0.00 & 0.02 \\
    Depression & 0.06 & 1 & 0.06 & 1 & 0.02 & 0.55 & 0.03 & 0.53 \\
    Disability & 0.00 & 0.09 & 0.00 & 0.12 & 0.01 & 0.52 & 0.01 & 0.57 \\
    Anxiety  & -0.00 & 0.03 & -0.00 & 0.02 & 0.00 & 0.02 & 0.00 & 0.02 \\
    Age &  & & 0.12 & 1 & & & 0.00 & 0.09 \\
    Sex &  & & -0.61 & 1 & & & -0.44 & 0.97 \\
    \midrule
    ROC AUC & \multicolumn{2}{c}{0.64} & \multicolumn{2}{c}{0.72} & \multicolumn{2}{c}{0.58} & \multicolumn{2}{c}{0.60} \\
    \bottomrule
  \end{tabular}
\end{table}

\begin{table}[!htbp]
  \caption{Stepwise and Lasso results using $y= \text{baseline NEET status}$}
  \label{tab:stepwise}
  \centering
  \begin{tabular}{l S[table-format=1.2] S[table-format=1.2] S[table-format=1.2] S[table-format=1.2] S[table-format=1.2] S[table-format=1.2] S[table-format=1.2] S[table-format=1.2] S[table-format=1.2] S[table-format=1.2] S[table-format=1.2] S[table-format=1.2]}
    \toprule
    & \multicolumn{6}{c}{(a) Stepwise results} & \multicolumn{6}{c}{(b) Lasso results}\\
    \cmidrule{2-13}
    & \multicolumn{3}{c}{(i) Ex. age, sex} & \multicolumn{3}{c}{(ii) Inc. age, sex} & \multicolumn{3}{c}{(i) Ex. age, sex} & \multicolumn{3}{c}{(ii) Inc. age, sex} \\
    \cmidrule{2-13}
    Variable & {$\hat{\beta}$} & {SE} & {$p$} & {$\hat{\beta}$} & {SE} & {$p$} & {$\hat{\beta}$} & {SE} & {$p$} & {$\hat{\beta}$} & {SE} & {$p$} \\
    \midrule
    Alcohol & & & & -0.03  & 0.01 & 0.01 & -0.10 & 0.08 & 0.21 & -0.27 & 0.15 & 0.08 \\
    Cannabis   & 0.02  & 0.01 & 0.02 & 0.02  & 0.01 & 0.02 & 0.13 & 0.08 & 0.07 & 0.22 & 0.14 & 0.11 \\
    Tobacco & & & & & & &  0.03 & 0.16 & 0.85 \\
    Depression & 0.05  & 0.01 & 0.00 & 0.06  & 0.01 & 0.00 & 0.11 & 0.10 & 0.27 & 0.23 & 0.17 & 0.17 \\
    Functioning & & & & & & & 0.12 & 0.10 & 0.22 & 0.23 & 0.17 & 0.18 \\
    Anxiety & & & & & & & & & & & & \\
    Age & & & & 0.12  & 0.03 & 0.00 & & & &  0.16 & 0.14 & 0.22 \\
    Sex & & & & -0.58  & 0.13 & 0.00 & & & & -0.32 & 0.13 & 0.01 \\
    \bottomrule
  \end{tabular}
\end{table}

\small

\clearpage

\bibliography{nips_2016}

\end{document}